\documentclass[twocolumn,showpacs,preprintnumbers,amsmath,amssymb,aps]{revtex4}

\usepackage{graphicx}              
\usepackage{color}
\usepackage{amssymb}

\newcommand{\ket}[1]{\left\vert#1\right\rangle}

\newcommand{\sandwich}[3]{\left< #1 \vphantom{#2 #3} \right| #2 \left|\vphantom{#1 #2} #3 \right>}

\newcommand{\numop}[1]{\displaystyle #1^{\dagger} #1}

\newcommand{\prim}{^{^{\prime}}}

\begin{document}

\title{Superconducting Spin Qubits}

\author{C. Padurariu}
\author{Yu. V. Nazarov}
\affiliation{Kavli Institute of NanoScience, Delft University of Technology, Lorentzweg 1, 2628 CJ, Delft, The Netherlands.}
\date{\today}


\begin{abstract}
We propose and theoretically investigate spin superconducting qubits. 
Spin superconducting qubit consists of a single spin confined in a Josephson junction. We show that owing to spin-orbit interaction, superconducting difference across the junction can polarize this spin. We demonstrate that this enables single qubit operations and more complicated quantum gates, where spins of different qubits interact via a mutual inductance of superconducting loop where the junctions are embedded.
Recent experimental realizations of Josephson junctions made of semiconductor quantum dots in contact with superconducting leads have shown that the number of electrons in the quantum dot can be tuned by a gate voltage. Spin superconducting qubit is realized when the number of electrons is odd.
We discuss the qubit properties at phenomenological level. We present a microscopic theory that enables us to make accurate estimations of the qubit parameters by evaluating the spin-dependent Josephson energy in the framework of fourth-order perturbation theory.
\end{abstract}

\maketitle


\pagenumbering{arabic}

\section{Introduction}

Potential benefits of a quantum computer: secure communication, fast 
database searching, efficient prime factorization \cite{Nie2000, Gro1997, Sho1997} 
have inspired significant research efforts. Building a quantum computer requires 
the realization of qubits as its elementary units. 
Useful qubits should satisfy two conditions: they can be manipulated and read 
before the quantum information stored in their state is lost, 
and they allow for engineering of a controllable interaction  between them. 
Designing and realizing such qubits defines the focus of most research in the field.  

Since the electron spin provides the simplest example of a coherent 
two-level system, that is, a qubit, spin-based qubits very soon became a subject
of intense theoretical \cite{Los1998} and experimental research. 
Experiments proved relatively long $T_1$ and $T_2$ times for single electron spins
trapped in quantum dots \cite{Han2003, Pet2005, Kop2007, Kop2008, Ama2008}, diamond 
\cite{Chi2006, Hans2008, Neu2008, Cap2009} and other materials.  
Good isolation from the environment protects from decoherence 
at cost of hampering qubit control and read-out. While current research 
successfully addresses these shortcomings \cite{Elz2004, Kop2006, Ata2006, Ber2008}, 
the realization of controllable interaction between pairs of spin qubits 
has been so far obstructed by numerous practical problems \cite{Han2008}. 

Superconducting qubits do better in this respect. 
Quantum logic gates involving the controllable interaction of two qubits 
have been demonstrated in a variety of setups \cite{Pas2003, Yam2003, Mcd2005, Ste2006, Pla2007}.  
Superconducting qubits exploit Josephson effect and Coulomb blockade,
both immediately related to electric variables of flux and charge.
This allows for easy integration of qubits into electric circuits 
and is the reason of the better performance, sensitivity to external 
noise that results in relatively short decoherence times \cite{Ste2006, Chi2003, Mar2005}.



\begin{figure}[h]
 \includegraphics[scale=0.26]{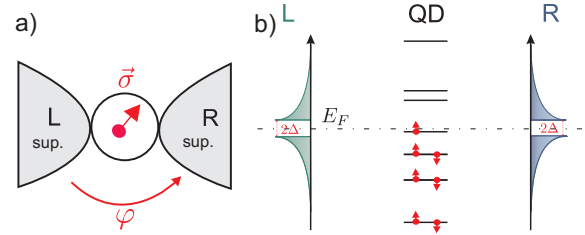}
 \caption{ Spin superconducting qubit. a) Quantum dot with odd number of electrons is connected to superconducting leads (L and R) biased at phase difference $\varphi$. b) Energy levels in quantum dot and the leads.
 \label{sys}}
 \end{figure}

In this article we propose a design of {\it superconducting spin qubit}
and discuss its feasibility and advantages. A spin is trapped in a quantum
dot connected to superconducting leads, thereby forming a Josephson junction, see Figure \ref{sys}.
Owing to spin-orbit interaction, the superconducting phase difference polarizes the spin. This 
provides means to read and manipulate its quantum state. 
We detail the operation of a single qubit and the design of qubit-qubit interaction 
that allows to make quantum logic gates. In particular, we emphasize the prospect of all 
electrical manipulation of the qubit state and qubit-qubit coupling. 
To prove feasibility of the design, we present microscopic calculations and 
numerical estimates of the spin- and gate voltage-dependent Josephson energy.

Quantum dot connected to superconducting leads is an essential element of our design.
Superconducting quantum dots have received theoretical attention rather early \cite{Ben1992, Khl1993}. 
About ten years later, experimental breakthrough has been achieved by making good contacts between 
semiconducting nanowires and superconducting leads \cite{Doh2005, Dam2006}. Gate electrodes put close to the nanowire can be used to create potential barriers in the nanowire, thereby defining a quantum dot.
The quantum dot is in the Coulomb blockade regime, 
that is, the number of electrons is tunable. It has been proven that such a dot 
can be included into a superconducting circuit as a Josephson junction carrying
a supercurrent \cite{Dam2006}.  
The idea of our proposal is to use this setup by keeping \textit{odd} number 
of electrons in the dot. In this case, the ground state of the resulting
Josephson junction is a spin doublet. This is advantageous in comparison to an earlier proposal
concerning Andreev quantum dot \cite{Cht2003}, 
where the spin doublet corresponds to an excited state of the system. Spin-orbit effects in quantum constrictions and dots have been discussed in \cite{Ben2006, Zaz2009}.

The structure of the article is as follows. In section \ref{phen}, we describe the qubit phenomenologically, discussing single qubit manipulation and design of qubit-qubit interaction. Section \ref{micro} focuses on the microscopic description of the Josephson junction with a single spin, leading to expressions of the spin-dependent and spin-independent parts of the Josephson energy in terms of the junction parameters. Section \ref{estim} discusses different parameter regimes, providing order of magnitude estimations of the Josephson energy. Section \ref{num} presents numerical calculations of the relevant quantities. Section \ref{conc} concludes.

\section{Phenomenological Description}
\label{phen}

We proceed with the phenomenological description of the qubit, postponing for now 
microscopic analysis. An important feature is that the superconducting current 
flowing through the Josephson junction is determined by the state of the spin enclosed. 
This can be understood as follows. 
Junction current is the result of transfer of Cooper pairs between the leads via 
coherent tunneling events.
Conventionally, tunneling events are spin conserving.
However, strong spin-orbit coupling mixes the spin states of electrons as they tunnel between 
orbitals in the superconductor and orbitals in the dot \cite{Dan2009}. 
The resulting non-spin-conserving tunneling amplitudes depend strongly on the
wavefunction of the levels involved. 
When a Cooper pair tunnels via two different levels in the dot, the initial spin-singlet 
configuration acquires complex spin structure, 
due to the different spin-dependent tunneling amplitudes.
If one of the two levels involved is occupied by a single electron, 
its spin restricts the pathways of Cooper pair tunneling,
due to Pauli exclusion. This directly couples current to spin, lifting
the degeneracy between spin up and spin down states.

\subsection{Effective Hamiltonian}

We describe the spin polarization effect by
the following junction Hamiltonian:
\begin{eqnarray}
H & = & E_j\ \cos\left(\varphi\right) + \left(\vec{\epsilon}_{\rm so}\:\cdot\:\vec{\sigma}\right)\ \sin\left(\varphi\right)\quad, 
\label{1}
\end{eqnarray}
where $\vec{\sigma}$ is the spin operator. The pseudovector $\vec{\epsilon}_{\rm so}$ 
defines the polarization axis in three dimensions. Its magnitude and direction depends on the spin-dependent tunneling amplitudes as well as on the positions of the levels in the quantum dot.
As a result, $\vec{\epsilon}_{\rm so}$ is independent of the superconducting phase 
difference $\varphi$, but does depend on gate voltage, as the gate electric field
modifies quantum dot wavefunctions and levels.

We will present estimations of $E_j$ and $\vec{\epsilon}_{\rm so}$ in section \ref{estim}. For present purposes it is enough to assume that typical values of $\vec{\epsilon}_{\rm so}$ are somewhat smaller than $E_j$. Actually, $E_j$ can be made zero by a certain choice of gate voltages. In the vicinity of this point, $\vec{\epsilon}_{\rm so}$ may be bigger than $E_j$. However, we do not concentrate on this case.

\begin{figure}[h]
 \includegraphics[scale=0.35]{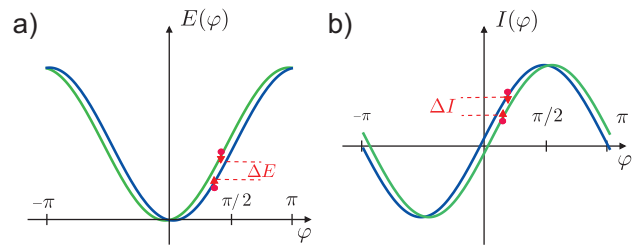}
 \caption{ Two states of a simple superconducting circuit (Figure \ref{circ}a) containing the qubit junction. a) Energy as a function of phase drop over the qubit junction $\varphi$. b) Loop current $I(\varphi)$.
 \label{spec}}
 \end{figure}

The junction forms a qubit: it may be found in two spin states, $\ket{\uparrow}$ and $\ket{\downarrow}$, that differ in spin projection along the polarization axis $\vec{\epsilon}_{\rm so}$.
At fixed phase, the energy spacing: 
$\Delta E = 2\:\left|\vec{\epsilon}_{\rm so}\right|\: \left|\sin\left(\varphi\right)\right|$,
leads to a measurable difference in superconducting current (Figure \ref{spec}):
\begin{eqnarray}
\Delta I & = & \pm
(2e/\hbar)\ \left|\vec{\epsilon}_{\rm so}\right|\: \left|\cos\left(\varphi\right)\right|\quad. 
\end{eqnarray}
The sign is determined by the direction of the spin along $\vec{\epsilon}_{\rm so}$. 
This provides the means to read the qubit state.

\begin{figure}[h]
 \includegraphics[scale=0.28]{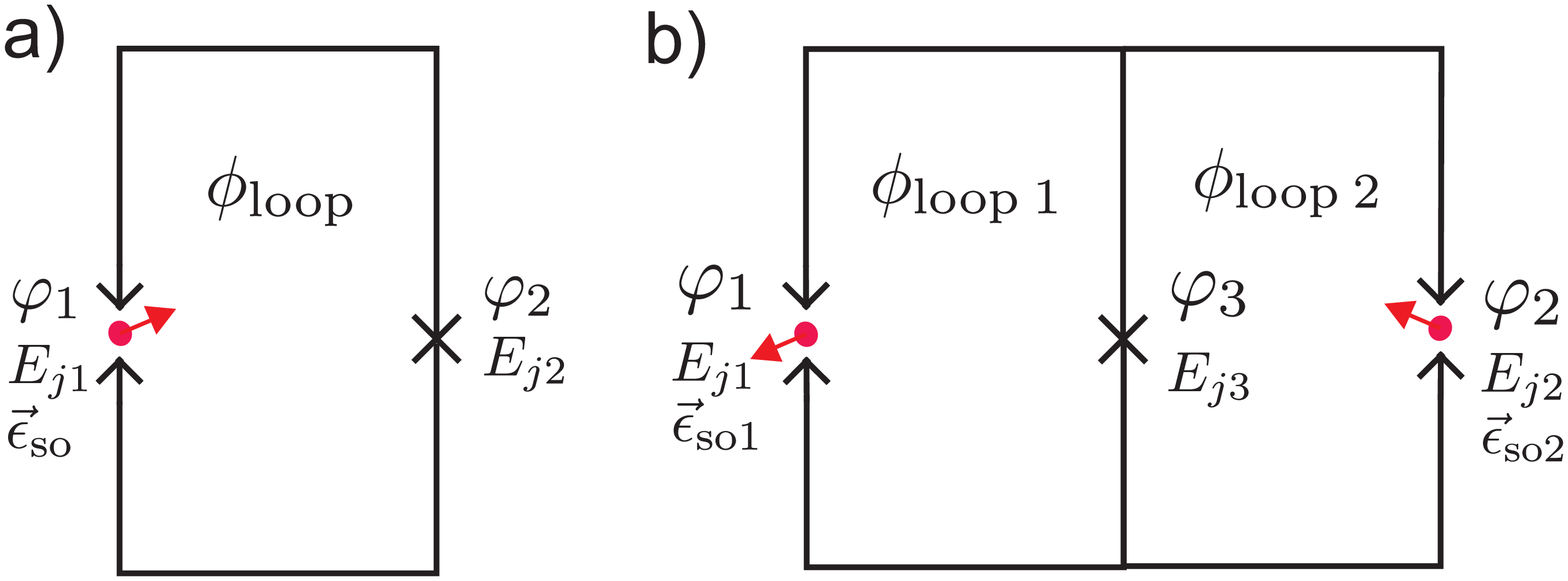}
 \caption{ Two simple superconducting circuits. a) Superconducting loop interrupted by qubit junction (characterized by $E_{j1}$ and $\vec{\epsilon}_{\rm so}$) and conventional Josephson junction (characterized by $E_{j2}$). The phase of the two junctions are modulated by the magnetic flux: $\varphi_1+\varphi_2=2\pi\phi_{\rm loop}/\phi_0$. b) Two superconducting loops are presented, each interrupted by a qubit junction. The loops have a common side that is interrupted by a conventional Josephson junction $E_{j3}$. Magnetic flux flowing trough the loops is represented by $\phi_{\rm loop\ 1}$ and $\phi_{\rm loop\ 2}$.
 \label{circ}}
 \end{figure}

To illustrate, consider a simple superconducting circuit consisting of a loop interrupted by
the qubit junction and a conventional Josephson junction, see Fig. \ref{circ}a. We apply magnetic flux through the loop $\phi_{\rm loop}$.
When the Josephson energy of the conventional junction is much larger than the Josephson energy of the qubit $E_{j2}\gg E_{j1}$, the phase induced by the magnetic flux is acquired mainly by the qubit junction $\varphi_2\approx 0$. The phase difference over the qubit junction can be fixed by fixing the magnetic flux $\varphi_1\approx 2\pi\phi_{\rm loop}/\phi_0$, $\phi_0=\pi\hbar/e$ being the magnetic flux quantum. The current through the loop is plotted as a function of phase for each of the two states of the qubit (see Figure \ref{spec}). The magnetic flux generated by this current can be measured by a nearby SQUID loop, not shown in Figure \ref{circ}.
This is a common technique used to measure the state of superconducting qubits \cite{Chi2003}.

The maximum value of the induced flux is achieved when $E_{j1}\approx E_{j2} \approx \left|\vec{\epsilon}_{\rm so}\right|$ and is of the order of the magnetic flux quantum $\phi_0$.
It is interesting to compare the induced flux, with the flux generated
by the magnetic dipole moment of the electron spin confined in the quantum dot. Considering an area element perpendicular to the dipole moment and situated at the distance $r$ from it, the flux flowing through the area scales with the distance as $\phi\propto r^{-1}$. We find the distance $r_0$ where the magnetic field of a single electron produces a quantum of magnetic flux via the relation $e^2/(4\pi \epsilon_0 r_0) \approx m_e c^2$. This distance is well into the subatomic region, in the order of $r_0 \approx 10^{-15}$ m.

\subsection{Spontaneous Currents}

Let us consider another regime, which is useless for qubit applications, but interesting from the point of view of general physics. Let us consider vanishing flux through the loop and assume formula $E_{j1}, E_{j2}\ll \left|\vec{\epsilon}_{\rm so}\right|$. In this case, the phase dependent energy Eq. (\ref{1}) can be expanded at small $\varphi$ 
\[
E(\varphi) = \left(E_{j1}+E_{j2}\right) \varphi^2/2 \pm \left|\vec{\epsilon}_{\rm so}\right| \varphi\quad ,
\]
with opposite signs corresponding to opposite spin orientations.
 \begin{figure}[h]
 \includegraphics[scale=0.33]{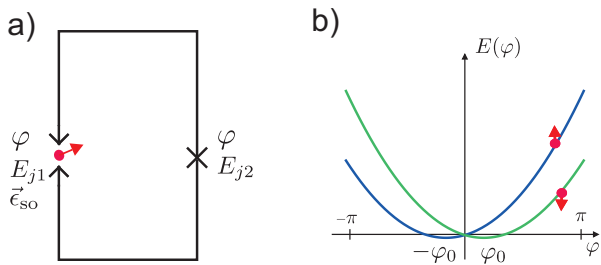}
 \caption{ Superconducting loop without magnetic flux. The dependence of energy on phase shows that the minimum energy is achieved at non-vanishing phase $\pm\varphi_0$. The sign of the equilibrium phase depends on the orientation of the spin in the qubit junction.
 \label{ano}}
 \end{figure}
 
We see (Figure \ref{ano}) that the equilibrium superconducting phase is non-zero. It takes opposite values depending on spin orientation. The current flowing through the junction is also non-zero $I = E_{j2}\:\varphi_0$. We stress that at zero flux applied, the system is time-reversible. The situation just described can be envisaged as spontaneous breaking of time reversibility. Indeed, the energies of the states with opposite spins are precisely the same. We will present detailed description of the situation elsewhere.

\subsection{Single Qubit Manipulation}

Let us turn our attention to manipulation of the qubit state. 
It is common to use pulses of an ac field of resonant frequency
$\omega=2\:\left|\vec{\epsilon}_{\rm so}\right|/\hbar\: \left|\sin\left(\varphi\right)\right|$.
Magnetic fields perpendicular to the loop plane induce modulations of superconducting phase, but
do not allow resonant manipulation as this does not change $\vec{\epsilon}_{\rm so}$.

In-plane magnetic fields can polarize qubit spin, deflecting it from the direction of $\vec{\epsilon}_{\rm so}$. 
However, this approach 
is difficult to realize experimentally due to misalignment. 
The magnetic field component perpendicular to the loop contributes to the total magnetic flux, 
changing the properties of the qubit at the same time as the manipulation is performed. 


Fortunately, magnetic field is not needed.
Rabi oscillations can also be induced electrically via the gate electrodes. We remind that the direction of $\vec{\epsilon}_{\rm so}$ depends on the position of levels in the dot. Therefore, shifting the gate voltage would also rotate $\vec{\epsilon}_{\rm so}$.
To illustrate, let us assume that a change of the gate voltage leads to a
corresponding change in polarization pseudovector 
$\vec{\epsilon}_{\rm so}\rightarrow\vec{\epsilon}_{\rm so}+\delta\vec{\epsilon}_{\rm so}$.
We describe the effect of the resonant pulse 
by the time-dependent qubit Hamiltonian:
\begin{eqnarray}
H(t) & = & \left[\left|\vec{\epsilon}_{\rm so}\right|\:\sigma_z+\delta\vec{\epsilon}_{{\rm so}}\cdot\vec{\sigma}\ \cos\left(\omega t\right)\right]\ \sin\left(\varphi\right)\quad,
\end{eqnarray}
where the $z$-axis is chosen along $\vec{\epsilon}_{\rm so}$.
Assuming $\left|\delta\vec{\epsilon}_{\rm so}\right|\ll\left|\vec{\epsilon}_{\rm so}\right|$, 
the dynamics of the qubit is described by Rabi oscillations with frequency:
\begin{eqnarray*}
\hbar\Omega_{\rm R} = \sqrt{\left(\left|\vec{\epsilon}_{\rm so}\right|\right)^2+\left(\delta\epsilon_{{\rm so},x}\right)^2+\left(\delta\epsilon_{{\rm so},y}\right)^2}\:|\sin\left(\varphi\right)|. 
\end{eqnarray*}

We stress that if the change $\delta\vec{\epsilon}_{\rm so}$ is parallel to the initial direction of $\vec{\epsilon}_{\rm so}$ Rabi oscillations do not occur. In this case the qubit retains its initial state. Thus, change of $\delta\vec{\epsilon}_{\rm so}$ in the perpendicular direction is essential for qubit manipulation.

We also mention that since electric fields are easier to localize in space than magnetic fields,
electrical manipulation is also advantageous in view of controlling 
qubits individually.

\subsection{Design of Qubit-Qubit Interaction}

Further, let us focus on qubit-qubit interaction. 
Two qubits can be included in a superconducting circuit, such that
the spin-dependent supercurrents flowing through the two qubits interact magnetically.
The interaction does not modify the polarization pseudovectors.
As a result, the qubit-qubit interaction is of the Ising type: 
\begin{eqnarray}
H					 & = &	H_1\:\displaystyle\sigma^z_1\:+\:
                  H_2\:\displaystyle\sigma^z_2\:+\: 
                  H_{12}\:\displaystyle\sigma^z_1\displaystyle\sigma^z_2\: ,
\end{eqnarray}
choosing the $z$-axis along $\vec{\epsilon}_{\rm so}$ for each qubit.
The Ising-type interaction is sufficient to perform the CNOT operation, 
which, in combination with single qubit operations,
enables universal quantum computations \cite{Eke1996}.

The CNOT gate is an operation on two qubits which has 
the effect of changing the state of one (target) qubit,
only when the other (control) qubit is in the excited state.
We propose a realization of CNOT gate using non-oscillating 
pulses of $H_{12}/\hbar$ of length $\tau$.
The pulse shifts the relative phase between two qubit states, as follows: 
states $\ket{\uparrow\uparrow}$ and $\ket{\downarrow\downarrow}$ gain 
phase factor $\exp(iH_{12}\tau/\hbar)$, while states 
$\ket{\uparrow\downarrow}$ and $\ket{\downarrow\uparrow}$ gain
phase factor $\exp(-iH_{12}\tau/\hbar)$).
Tuning the length of the pulse such that $H_{12}\tau/\hbar = \pi/4$,
one obtains a phase-shift gate ${\rm G}_{\rm phase}$ that can be combined with 
Bloch sphere rotations by $\pi/2$ of the control and target qubit 
around the coordinate axes ${\rm R}_{x,y,z}(\pi/2)$, to achieve the controlled-NOT gate:
\begin{eqnarray}
{\rm CNOT} & = & e^{i\pi/4}\ {\rm R}^{(1)}_z(\pi/2) {\rm R}^{(2)}_z(-\pi/2) \nonumber\\
\ & \ & \quad {\rm R}^{(2)}_x(\pi/2) {\rm G}_{\rm phase} {\rm R}^{(2)}_y(\pi/2)\ ,
\end{eqnarray}
where the superscripts $(1)$ and $(2)$ imply that the operation is performed 
on the control and target qubit, respectively.

We have the opportunity to tune the interaction and also to switch it on and off.
To give an example, consider the following circuit: two qubits 
with their superconducting loops connected in parallel (Fig. \ref{circ}b).
The interaction energy can be tuned by the following parameters:
the magnetic flux flowing through the superconducting loops 
($\phi_{{\rm loop}\:1}$ and $\phi_{{\rm loop}\:2}$) and 
the Josephson energy of the conventional Josephson junction $E_{j3}$.
To lowest order in the polarization pseudovectors $\left|\vec{\epsilon}_{\rm so 1,2}\right|$, 
the interaction takes the form:
\begin{eqnarray}
H_{12} & = & - \frac{\left|\vec{\epsilon}_{\rm so 1}\right|\: \left|\vec{\epsilon}_{\rm so 2}\right|}{|\tilde{E}|}\ \cos\left(\varphi_{\tilde{E}}+\varphi_{1}\right)
\cos\left(\varphi_{\tilde{E}}+\varphi_{2}\right) ,
\end{eqnarray}
where the complex-valued energy $\tilde{E}$ denotes:
$\tilde{E} = E_{j1}e^{i \varphi_{1}}+E_{j2}e^{i \varphi_{2}}+E_{j3}$, 
and $\varphi_{\tilde{E}}$ denotes its complex argument 
$\varphi_{\tilde{E}} = \tan^{-1}\left[{\rm Im}\: \displaystyle\tilde{E}/{\rm Re}\: \displaystyle\tilde{E}\right]$.
The angles $\varphi_{1,2}$ are proportional to the magnetic fluxes: 
$\varphi_{1,2} = 2\pi\left(\displaystyle\phi_{{\rm loop}\:1,2}/\displaystyle\phi_0\right)$, 
with $\phi_0$ the magnetic flux quantum. 
Setting $\varphi_1=\varphi_2=\pi/2$, it is possible to tune 
the interaction simply by controlling $E_{j3}$. 
For $E_{j3}\gg(E_{j1}+E_{j2})$ the interaction is turned off $H_{12}\rightarrow 0$, 
while in the opposite limit $E_{j3}\ll(E_{j1}+E_{j2})$ the interaction achieves 
a maximum $H_{12}=- \left|\vec{\epsilon}_{\rm so 1}\right|\: \left|\vec{\epsilon}_{\rm so 2}\right|/|\tilde{E}|$. 
Such control of the conventional Josephson energy can be achieved 
using exclusively electrical means, as was demonstrated in \cite{Dam2006}.

\section{Microscopic Description}
\label{micro}
\subsection{Hamiltonian}

The total Hamiltonian comprises terms describing leads (left and right), quantum dot and tunneling between electronic states in the leads and in the dot:
\begin{eqnarray}
\hat{H} & = & \hat{H}_{\rm L} + \hat{H}_{\rm R} + \hat{H}_{\rm QD} + \hat{H}_{\rm T}\ ,
\label{Htot}\\
\ & = & \hat{H}_0 + \hat{H}_{\rm T} \ .\nonumber
\end{eqnarray}
The electronic states of the left lead are labeled by $l$ and spin index $\sigma$ and are affected by the superconducting order parameter $\Delta e^{i\varphi_L}$.
\begin{eqnarray}
\hat{H}_{\rm L} & = & \displaystyle\sum_{l,\sigma} \xi_l\ \numop{a_{l\sigma}} + \nonumber \\
\ & \ & \Delta e^{i\varphi_L} g^{\sigma\sigma\prim} a^{\dagger}_{l\sigma\prim}a^{\dagger}_{l\sigma} + g^{\sigma\sigma\prim}\Delta e^{-i\varphi_L} a_{l\sigma}a_{l\sigma\prim}\ ,
\label{Hel}							 
\end{eqnarray}
were $a^{\dagger}$ ($a$) denotes the creation (annihilation) operators and $\xi$ denotes the energy of the levels in the normal metal state, counted from the chemical potential of the lead.
The matrix $g=i\sigma^2$ ensures that the two electrons are a spin singlet when forming a Cooper pair in the leads. In the superconducting state, the energy of the levels is $\epsilon_l = \sqrt{\xi_l^2+\Delta^2}$.

Similarly, electronic states in the right lead are labeled by $r$ and are affected by $\Delta e^{i\varphi_R}$.

The Hamiltonian of the quantum dot contains charging energy term along with the terms describing non-interacting electrons in levels labeled by $m$, $\displaystyle\sum_{m} \xi_m \numop{a_m}$,
\begin{eqnarray}
\hat{H}_{\rm QD} & = & \displaystyle\sum_{m} \xi_m \numop{a_m} + \left(\hat{N}-N_0\right)^2 E_C\ ,
\label{Hq}
\end{eqnarray}
with $\hat{N}$ being the operator of the total number of electrons $\hat{N}=\displaystyle\sum_{m}\numop{a_m}$. $N_0= V_gC_g/e$ represents the effect of gate voltage which allows to tune the total number of electrons in the dot. $C_g$ here is the capacitance to the gate. Since it only appears in combination with gate voltage, we find it convenient to rescale the gate voltage $V_g\: C_g/C \rightarrow V_g $, with $C$ being the total capacitance of the junction. In these units, $N_0=eV_g/E_C$.

We stress that all these terms are not affected by spin-orbit coupling. The reason for this is that spin-orbit coupling retains the double degeneracy of the electronic states. Instead of spin doublets they become Kramers doublets, so index $\sigma$ in this case refers to the components of Kramers doublet, rather than to the original spin state.
The only place where spin-orbit coupling plays a role is the tunneling part of the Hamiltonian and can be included into spin-dependent tunneling amplitudes \cite{Dan2009}:
\begin{eqnarray}
\hat{H}_{\rm T} & = & \displaystyle\sum_{l,r,m,\rho,\sigma} T^{\rho\sigma}_{lm} a^{\dagger}_{m\sigma}a_{l\rho} + T^{\rho\sigma}_{mr} a^{\dagger}_{r\sigma}a_{m\rho} + {\rm H. c.}\ ,
\label{Ht}
\end{eqnarray}
where $\rho$ and $\sigma$ are spin indexes.\\
As a result of time reversal symmetry, we may write the tunneling matrices in spin space:
\begin{eqnarray}
T^{\rho\sigma}_{lm} & = & T^{0}_{lm}\ \delta_{\rho\sigma} + i\ \displaystyle\sum^{3}_{j=1} t^{j}_{lm} \sigma^{j}_{\rho\sigma} \ .
\label{T}
\end{eqnarray}
where $T^{0}_{lm}$ and $t^{j}_{lm}$ are chosen to be real. The spin-independent part is symmetric $T^{0}_{lm} = T^{0}_{ml}$ and spin-dependent part is antisymmetric $t^{j}_{lm}=-t^{j}_{ml}$ with respect to interchanging the state indices.

In principle the tunneling matrix elements can be evaluated by computing overlap of spin-dependent wavefunctions of corresponding components of Kramers doublets. It is known that wavefunctions become random either because of scattering at defects in the leads and in the dot, or because of chaoticity of classical electron motion in the dot. If we consider extended system, the statistics of spin-dependent part of such overlaps is governed by length scale $l_{\rm sf}$ that is a spin flip length, induced by spin-orbit scattering. At distances exceeding this scale, spin orientation of Kramers doublets becomes completely random, so that $T^0$ is of the order of $\left|\vec{t}\right|$. The length scale $l_{\rm sf}$ corresponds to time scale $\tau_{\rm sf}$, spin-flip time. If we apply this now to a quantum dot, the spin structure becomes random for each level in the dot, provided the mean level spacing in the dot $\delta_S$ is comparable or smaller than $\hbar/\tau_{\rm sf}$. The ratio of the two defines the strength of spin-orbit coupling. For our estimations, we assume $\vec{t}/T^0\approx 0.1$ and random distribution of the direction. Preliminary experiments with InAs nanowires \cite{Stevan} confirm this by order of magnitude.

\subsection{Josephson Energy}

In the following we provide a microscopic description of the Josephson effect in the presence of
spin-orbit coupling.
We treat Cooper pair transport through a quantum dot 
in the regime of Coulomb blockade, with no bias voltage 
and disregarding effects of thermal excitation. 
Within the Coulomb diamonds, the only mechanism of transport is cotunneling.  
Four tunneling events are sufficient to transfer a Cooper pair between the leads \cite{Bau1994}
and these processes dominate for sufficiently small normal state conductances $G_N \ll G_Q\equiv e^2/\pi\hbar$. 
We employ fourth-order perturbation theory in the tunneling amplitudes
and find the Josephson energy as the perturbation correction to the ground state energy.

The ground state of the unperturbed Hamiltonian $\hat{H}_0=\hat{H}_{\rm L} + \hat{H}_{\rm R} + \hat{H}_{\rm QD}$ is degenerate with respect to the number of Cooper pairs on the superconducting leads. The wave function can be written as a direct product of three wave functions: $\psi\left(N_{\rm L},N_{\rm R}\right)=\psi_{\rm L}\left(N_{\rm L}\right)\otimes\psi_{\rm QD}\otimes\psi_{\rm R}\left(N_{\rm L}\right)$ corresponding to the two leads and the quantum dot. $N_{\rm L}$ and $N_{\rm R}$ represent the number of electrons in the respective leads.

In the presence of tunneling, the degeneracy of the ground state is lifted and the energy splitting between states with different number of Cooper pairs in the leads corresponds to the Josephson energy:
\begin{eqnarray}
E_J(\varphi) & = & \sandwich{\psi\left(-2,2\right)}{\hat{T}^{(4)}}{\psi\left(0,0\right)}e^{i\varphi} + \nonumber\\
\         & \ & \sandwich{\psi\left(2,-2\right)}{\hat{T}^{(4)}}{\psi\left(0,0\right)}e^{-i\varphi} \nonumber\\
\         & = & 2\: {\rm Re}\left\{\sandwich{\psi\left(-2,2\right)}{\hat{H}^{(4)}_{\rm T}}{\psi\left(0,0\right)}e^{i\varphi}\right\}
\label{Ejff}
\end{eqnarray}
where $N_{\rm L}$ and $N_{\rm R}$ have been set to zero. 

Here, operator $\hat{T}^{(4)}$ represents the correction to the amplitude of the fourth order in tunneling Hamiltonian:
\begin{eqnarray}
\hat{T}^{(4)} = \hat{H}_{\rm T}\frac{1}{E_0-\hat{H}_0}\hat{H}_{\rm T}\frac{1}{E_0-\hat{H}_0}\hat{H}_{\rm T}\frac{1}{E_0-\hat{H}_0}\hat{H}_{\rm T} \ .
\label{Ht4}
\end{eqnarray}

The structure of the fourth-order correction is as follows. Each tunneling operator $\hat{H}_{\rm T}$ describes 
the transition of one electron between a state in the dot and a state 
in one of the superconducting leads. 
There are 24 distinct sequences resulting from the permutations of the four tunneling 
events considered. Each process involves three intermediary states. The energies of the virtual states appear as the three denominators of Eq. \ref{Ht4}. 
The three energy denominators unambiguously characterizes the sequence of tunneling events.\\
In contrast, the spin structure does not depend on the order of individual tunneling events. 
The spin-structure of $\hat{T}^{(4)}$ can be recovered from the product of four 
tunneling amplitudes describing hopping of the two electrons between the leads and the quantum dot:
\begin{multline}
\displaystyle\sum_{l,r} g^{\rm T}\ T^{\rm T}_{lm}\ g\ T_{ln} T_{nr}\ g^{\rm T}\ T^{\rm T}_{mr}\ g\;\delta\left(\xi_L-\xi_l\right)\;\delta\left(\xi_R-\xi_r\right) = \\
\shoveleft{\displaystyle\sum_{l,r} T_{ml}T_{ln}T_{nr}T_{rm}\;\delta\left(\xi_L-\xi_l\right)\;\delta\left(\xi_R-\xi_r\right)} = \\ 
\shoveleft{P^0_{mn}\left(\xi_L, \xi_R\right) + i \vec{P}_{mn}\left(\xi_L, \xi_R\right)\cdot\vec{\sigma}\qquad\qquad,}
\label{polari}
\end{multline}
where $m$ and $n$ label states of the quantum dot.

Let us explain the properties of the two terms in Eq. (\ref{polari}). 
In the following, we neglect terms proportional to the square of the small term $\vec{t}/T^0$.
Since the leads have continuous spectra, it is convenient to introduce transport rates 
which are continuous functions of energy:
$
\Gamma_{L(R),m}(\xi) = 2\pi/\hbar\: \displaystyle\sum_{l(r)} \left|\displaystyle T^0_{ml(r)}\right|^2 
\delta(\xi-\xi_{l(r)})
$.
The spin-independent term $P^0_{mn}\left(\xi_L, \xi_R\right)$, has different properties in the case $m=n$, compared to the case $m\neq n$. Terms diagonal in the dot levels have the following simple form: $P^0_{mm}\left(\xi_L, \xi_R\right) = (\hbar/2\pi)^2\Gamma_{L,m}(\xi_L)\Gamma_{R,m}(\xi_R)$. As a result, they are always positive. In contrast, the sign of $P^0_{mn}\left(\xi_L, \xi_R\right)$ when $m\neq n$ can be related to the parities of the wavefunctions of the two dot states. In analogy to the case of a particle in a box, the parity of wavefunctions in a quantum dot alternates between neighboring states. As a result, the sign of $P^0_{m,n\neq m}\left(\xi_L, \xi_R\right)$ alternates and the sum over all states in the quantum dot averages out: $\displaystyle\sum_{m,n\neq m}P^0_{mn}\left(\xi_L, \xi_R\right)\approx 0$.

The contribution corresponding to the spin-dependent term in Eq. (\ref{polari}) changes sign for different spin orientations. Therefore, this term cancels out if the levels $m$ and $n$ are 
both either empty or filled with electrons, as there is no asymmetry between spin up and spin down terms. 
However, if either state $m$ or $n$ is the state filled with a single electron, 
the contribution is finite, giving rise to the spin polarization effect.

After calculating the energy denominators for each tunneling sequence and integrating over all states in the leads and all pairs of states in the dot, we obtain the following result for the spin-independent Josephson energy \cite{Roz2001}:
\begin{widetext}
\begin{eqnarray}
E_j & = & \frac{\Delta^2}{4}  \int^{\infty}_{-\infty}\frac{d\xi_L}{\epsilon_L}\int^{\infty}_{-\infty}\frac{d\xi_R}{\epsilon_R} \left[\displaystyle\sum_{M_1,M_2} P^0_{M_1M_2}\left(\xi_L, \xi_R\right) J_{ee}(\xi_L,\xi_R,\xi_{M_1},\xi_{M_2}) + \displaystyle\sum_{m_1,m_2} P^0_{m_1m_2}\left(\xi_L, \xi_R\right) J_{ee}(\xi_L,\xi_R,\xi_{m_1},\xi_{m_2})\right. -\nonumber\\
\             & \ & 2\displaystyle\sum_{M,m}\ P^0_{Mm}\left(\xi_L, \xi_R\right) J_{eh}(\xi_L,\xi_R,\xi_{M},\xi_{m}) + \displaystyle\sum_{M}\ P^0_{M0}\left(\xi_L, \xi_R\right) \left(J_{ee}\left(\xi_L,\xi_R,\xi_M,\xi_0\right) -J_{eh}\left(\xi_L,\xi_R, \xi_M,\xi_0 \right)\right) +\nonumber\\
\             & \ & \left.\displaystyle\sum_{m} P^0_{0m}\left(\xi_L, \xi_R\right) \left(J_{hh}\left(\xi_L,\xi_R,\xi_0,\xi_m\right) -J_{eh}\left(\xi_L,\xi_R, \xi_0,\xi_m \right)\right) - 2\displaystyle P^0_{00}\left(\xi_L, \xi_R\right) J_{eh}(\xi_L,\xi_R,\xi_{0},\xi_{0}) \right]\ ,
\label{Ej}
\end{eqnarray}
\end{widetext}
and for the polarization pseudovector:
\begin{widetext}
\begin{eqnarray}
\vec{\epsilon}_{\rm so} & = & \displaystyle\frac{\Delta^2}{4}\   \displaystyle\int^{\infty}_{-\infty}\displaystyle\frac{d\xi_L}{\epsilon_L}\;\displaystyle\int^{\infty}_{-\infty}\displaystyle\frac{d\xi_R}{\epsilon_R} \left[ \frac{}{} \displaystyle\sum_{M}\ \vec{P}_{M0}\left(\xi_L, \xi_R\right) \left(J_{ee}\left(\xi_L,\xi_R,\xi_M,\xi_0\right) +J_{eh}\left(\xi_L,\xi_R, \xi_M,\xi_0 \right) \right)-\right.\nonumber\\
\ & \ & \left.\displaystyle\sum_{m}\ \vec{P}_{0m}\left(\xi_L, \xi_R\right) \left(J_{hh}\left(\xi_L,\xi_R,\xi_0,\xi_m\right) +J_{eh}\left(\xi_L,\xi_R, \xi_0,\xi_m \right) \right)\right] \ ,
\label{epsso}
\end{eqnarray}
\end{widetext}
where $m$, $m_1$ and $m_2$ label filled dot states, $M$, $M_1$, $M_2$ empty dot states and $0$ labels the half-filled state. We have used the notation $\epsilon_{L,R}=\sqrt{\Delta^2+\xi^2_{L,R}}$. The $J$-functions contain the energy denominators and are different for processes where the Cooper pair is transfered via two electrons, two holes or one electron and one hole, respectively:
\begin{widetext}
\begin{multline*}
J_{ee}(\xi_L,\xi_R,\xi_m,\xi_n) = \frac{1}{\epsilon_L+\xi_m+E(e)}\frac{1}{\epsilon_L+\epsilon_R}\frac{1}{\epsilon_R+\xi_n+E(e)} + \frac{1}{\epsilon_L+\xi_n+E(e)}\frac{1}{\epsilon_L+\epsilon_R}\frac{1}{\epsilon_R+\xi_m+E(e)} +\\ \frac{1}{\epsilon_L+\xi_m+E(e)}\frac{1}{\xi_m+\xi_n+E(2e)}\frac{1}{\epsilon_R+\xi_m+E(e)} + \frac{1}{\epsilon_L+\xi_n+E(e)}\frac{1}{\xi_m+\xi_n+E(2e)}\frac{1}{\epsilon_R+\xi_m+E(e)} +\\ \frac{1}{\epsilon_L+\xi_m+E(e)}\frac{1}{\xi_m+\xi_n+E(2e)}\frac{1}{\epsilon_R+\xi_n+E(e)} +\frac{1}{\epsilon_L+\xi_n+E(e)}\frac{1}{\xi_m+\xi_n+E(2e)}\frac{1}{\epsilon_R+\xi_n+E(e)}\ ,\\
J_{hh}(\xi_L,\xi_R,\xi_m,\xi_n) = \frac{1}{\epsilon_L-\xi_m+E(-e)}\frac{1}{\epsilon_L+\epsilon_R}\frac{1}{\epsilon_R-\xi_n+E(-e)} + \frac{1}{\epsilon_L-\xi_n+E(-e)}\frac{1}{\epsilon_L+\epsilon_R}\frac{1}{\epsilon_R-\xi_m+E(-e)} +\\ \frac{1}{\epsilon_L-\xi_m+E(-e)}\frac{1}{-\xi_m-\xi_n+E(-2e)}\frac{1}{\epsilon_R-\xi_m+E(-e)} +\frac{1}{\epsilon_L-\xi_n+E(-e)}\frac{1}{-\xi_m-\xi_n+E(-2e)}\frac{1}{\epsilon_R-\xi_m+E(-e)} +\\ \frac{1}{\epsilon_L-\xi_m+E(-e)}\frac{1}{-\xi_m-\xi_n+E(-2e)}\frac{1}{\epsilon_R-\xi_n+E(-e)} +\frac{1}{\epsilon_L-\xi_n+E(-e)}\frac{1}{-\xi_m-\xi_n+E(-2e)}\frac{1}{\epsilon_R-\xi_n+E(-e)}\ ,\\
J_{eh}(\xi_L,\xi_R,\xi_m,\xi_n) =  \frac{1}{\epsilon_L+\xi_m+E(e)}\frac{1}{\epsilon_L+\epsilon_R}\frac{1}{\epsilon_L-\xi_n+E(-e)} +  \frac{1}{\epsilon_R+\xi_m+E(e)}\frac{1}{\epsilon_L+\epsilon_R}\frac{1}{\epsilon_R-\xi_n+E(-e)} +\\ \frac{1}{\epsilon_L+\xi_m+E(e)}\frac{1}{\xi_m-\xi_n+\epsilon_L+\epsilon_R}\frac{1}{\epsilon_R+\xi_m+E(e)} + \frac{1}{\epsilon_R-\xi_n+E(-e)}\frac{1}{\xi_m-\xi_n+\epsilon_L+\epsilon_R}\frac{1}{\epsilon_R+\xi_m+E(e)} +\\
 \frac{1}{\epsilon_L+\xi_m+E(e)}\frac{1}{\xi_m-\xi_n+\epsilon_L+\epsilon_R}\frac{1}{\epsilon_L-\xi_n+E(-e)} +\frac{1}{\epsilon_R-\xi_n+E(-e)}\frac{1}{\xi_m-\xi_n+\epsilon_L+\epsilon_R}\frac{1}{\epsilon_L-\xi_n+E(-e)}\ .
\end{multline*}
\end{widetext}

\section{Estimations}
\label{estim}

Let us estimate the typical magnitude of $\vec{\epsilon}_{\rm so}$ and compare it to the magnitude of spin-independent Josephson energy. We will see that the relative magnitudes as well as its absolute value cannot be just estimated by the typical strength of spin-orbit coupling $\left|\vec{t}\right|/T_0$. The estimations depend on three energy scales in the problem: $\delta_S$, $E_C$ and $\Delta$. Besides, it depends on the energy distance to the diamond edge.

It has been already shown in \cite{Bau1994} that Josephson energy and critical current exhibit spectacular peculiarities at the diamond edge. These peculiarities are no singularities. This is related to the diamond structure in the presence of superconductivity \cite{Bau1994}. There is a bistability region in interval of width $\Delta(e V_g) = 2\Delta$ around each diamond edge. The singularities in denominators of perturbation theory occur at the edges of bistability region for the charged state of higher energy. We always assume the dot to be in the lowest energy state. This saves us from singularities. However, the proximity to the point of singularity gives a spectacular increase or decrease in Josephson energy at the diamond edge. To account for this in our estimations we introduce the minimal energy distance to the edge of Coulomb diamond $E_{\pm} = |\min\left({E(e), E(-e)}\right)|>\Delta$. We envisage two separate situations: $E_{\pm}\approx E_C$ and $E_{\pm}\ll E_C$.

The dominant contribution to the Josephson energy consists of the terms with smallest energy denominators. The magnitude of the energy denominators depends on the energy of the dot levels involved in tunneling. We find that the energy interval for quantum dot levels involved in the dominating contribution has the width $E_{\pm}$, even in the regime $E_{\pm}\gg\Delta$.
Let us define the number of levels within this interval $N_S$ as the integer part of $E_{\pm}/\delta_S+1$. If $\delta_S\ge E_{\pm}$ the Josephson energy is dominated by the contribution of a single level. If $\delta_S< E_{\pm}$ there are multiple levels participating in the dominating contribution.

Using Eq. (\ref{Ej}) we find the following estimate for the spin-independent Josephson energy:
\[
E_j\ \approx\ N_S\:\frac{\Gamma_L\Gamma_R}{\Delta}\:\frac{\Delta^2}{E^2_{\pm}}\ .
\]
Similarly, we can estimate the spin-dependent component using Eq. (\ref{epsso}):
\[
\left|\vec{\epsilon}_{\rm so}\right|\ \approx\ \sqrt{N_S}\:\frac{\Gamma_L\Gamma_R}{\Delta}\ \frac{\Delta^2}{E_{\pm}\left(\max\left(E_{\pm},\delta_S\right)\right)}\ \frac{\left|\vec{t}\right|}{T_0}\ .
\]

In all parameter regimes considered, both $E_j$ as well as $\left|\vec{\epsilon}_{\rm so}\right|$ increase as $E_{\pm}/\Delta$ decreases, i.e. as we approach the edges of the Coulomb diamond. We can explain this in terms of the energy of the intermediary virtual states. In the middle of the Coulomb diamond the energy cost of adding an electron to the quantum dot is maximum and the high energy of the intermediary states reduces the probability of Cooper pair tunneling. Toward the edges of the diamond, tunneling processes involving the state closest to resonance will have intermediary states that are lower in energy, resulting in the increase in Josephson coupling.

Let us turn our attention to the multi-level regime, where Josephson tunneling is a result of the interference of tunneling processes that involve all pairs of the $N_S$ relevant states. The spin-dependent and spin-independent terms do not scale in the same way with the number of levels involved. In the case of $E_j$, the dominant contribution results from summation of diagonal spin-independent elements $P^0_{mm}\left(\xi_L, \xi_R\right)$. As a result, $E_j$ scales with the number of levels $N_S$. In the case of $\vec{\epsilon}_{\rm so}$, the result of summation over tunneling contributions is equivalent to the distance traveled in a random walk in three dimensions, after $N_S$ steps. The average result in this case scales as $\sqrt{N_S}$.

The ratio of spin-dependent to spin-independent contributions is:
\[
\frac{\left|\vec{\epsilon}_{\rm so}\right|}{E_j}\ \approx\ \frac{1}{\sqrt{N_S}}\ \frac{E_{\pm}}{\max\left(E_{\pm},\delta_S\right)}\ \frac{\left|\vec{t}\right|}{T_0}
\]
Apart from the trivial dependence on spin-orbit coupling strength, the ratio $\left|\vec{\epsilon}_{\rm so}\right|/E_j$ is further reduced by factors that depend on the gate voltage. We distinguish two important regimes: the single level regime $\delta_S\geq E_C$, and the multi-level regime $\delta_S\ll E_C$, where multiple dot states contribute.

In the single level regime we find $\left|\vec{\epsilon}_{\rm so}\right|/E_j \approx E_{\pm}/\delta_S$. Thus, it is large in the middle of the diamond, where it scales as $E_C/\delta_S$, decreasing towards the edge, where it scales as $\Delta/\delta_S$. The behavior has a simple explanation in terms of energy of intermediary virtual states. The dominant contribution to $E_j$ arises from processes involving tunneling of both Cooper pair electrons through the dot level closest to resonance. Such processes do not contribute to the polarization effect; as we have seen, only processes involving two different levels contribute. Thus, the energy of intermediary states decreases faster as we approach the diamond edge for dominant processes contributing to $E_j$.

We can also estimate the dependence of the direction of $\vec{\epsilon}_{\rm so}$ on the gate voltage. For this, we define the change of $\vec{\epsilon}_{\rm so}$ in the perpendicular direction, in the following way: $\left|\frac{d\vec{\epsilon}_{\rm so}}{d(eV_g)}\times\frac{\vec{\epsilon}_{\rm so}}{\left|\vec{\epsilon}_{\rm so}\right|}\right|$. This choice is motivated by the desire to achieve electrical manipulation of the qubit. We have shown in section \ref{phen} that change of $\vec{\epsilon}_{\rm so}$ perpendicular to its initial direction is an essential ingredient for observation of Rabi oscillations.

We find the following estimate of the perpendicular derivative:
\begin{eqnarray*}
\left|\frac{d\vec{\epsilon}_{\rm so}}{d(eV_g)}\times\frac{\vec{\epsilon}_{\rm so}}{\left|\vec{\epsilon}_{\rm so}\right|}\right|\ \approx
\frac{\Gamma_L\Gamma_R}{\Delta^2}\ \frac{\Delta^2}{E_{\pm} \left[\max\left(E_{\pm},\delta_S\right)\right]^2}\ \frac{\left|\vec{t}\right|}{T_0}\ .
\end{eqnarray*} 
To understand why the direction of $\vec{\epsilon}_{\rm so}$ depends on gate voltage, we study the structure of Eq. (\ref{epsso}). We have mentioned that $\vec{\epsilon}_{\rm so}$ is the sum of $N_S$ vectors characterizing spin-dependent tunneling via two different states in the dot, one being the half-filled state. The weight of each vector is determined by the energy denominators, and thus depends on gate voltage. It is thus necessary to include the contributions of at least three different levels, one being the half-filled state, in order to estimate the change of the orientation of $\vec{\epsilon}_{\rm so}$. As a result, for large level spacing the derivative in the perpendicular direction is reduced by a factor of $\left(E_{\pm}/\delta_S\right)^2$. In comparison, the derivative of the modulus of $\vec{\epsilon}_{\rm so}$ decreases slower for large level spacing, only as $E_{\pm}/\delta_S$.

\section{Numerical Analysis}
\label{num}

We perform numerical calculations of the spin-independent and spin-dependent parts of the Josephson energy, based on Eq. (\ref{Ej}) and (\ref{epsso}). We are particularly interested in the ratio $\left|\vec{\epsilon}_{\rm so}\right|/E_j$, relevant for experiments aiming to measure the separation between qubit states, and on the dependence of the direction of $\vec{\epsilon}_{\rm so}$ on gate voltage, relevant for experiments aiming to perform electrical manipulation of the qubit. 

To set up the calculations, we assume that the tunneling amplitudes are independent of the lead states. We account for the fact that experimental realization of quantum dots does not permit control of the resulting localized states: 
we choose random energy spacing between the dot states and random absolute value of tunneling amplitudes. The parities of the localized wavefunctions are also chosen randomly, resulting in a random sign associated to off-diagonal spin-independent contributions $P^0_{m,n\neq m}\left(\epsilon_L,\epsilon_R\right)$. The parameters characterizing the dot levels are the average level spacing $\delta_S$ and average modulus of the spin-independent and spin-dependent tunneling amplitudes, respectively $\langle\left|T^0\right|\rangle$ and $\langle\left|\vec{t}\right|\rangle$. As estimated, the ratio $\left|\vec{\epsilon}_{\rm so}\right|/E_j$ is proportional to $\langle\left|\vec{t}\right|\rangle/\langle\left|T^0\right|\rangle$ and we include this ratio in the energy unit of the spin-dependent term.

Additional parameters in the calculation are the superconducting energy gap $2\Delta$ and the charging energy $E_C$. For the numerical analysis, we need to consider a finite number of levels of the quantum dot. The results presented are obtained including a number of $N=20$ quantum dot levels.

\begin{figure}[h]
 \includegraphics[scale=0.27]{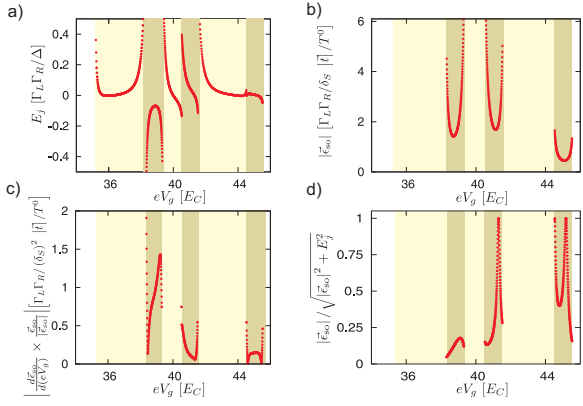}
 \caption{ Numerical results obtained for the regime: $E_C/\Delta=10$, $\delta_S/E_C=1.5$. Six diamonds are presented: diamonds with even number of electrons are represented by lighter shading, while darker shaded regions represent diamonds with odd number of electrons. For details regarding the quantities plotted, see Section \ref{num}.
 \label{f3}}
 \end{figure}

\begin{figure}[h]
 \includegraphics[scale=0.27]{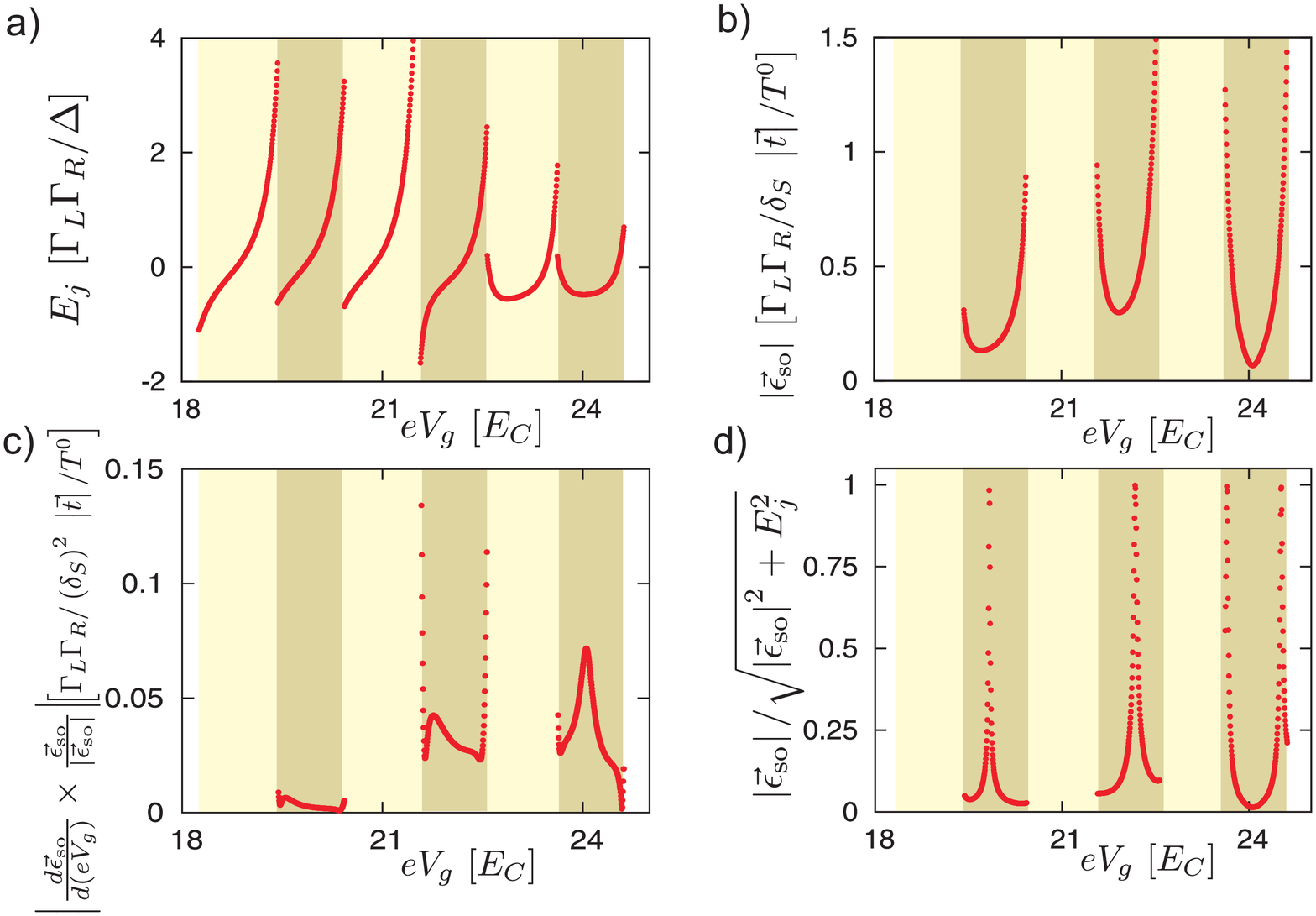}
 \caption{ The same as in Figure \ref{f3}, for the regime: $E_C/\Delta=10$, $\delta_S/E_C=0.1$.
 \label{f4}}
 \end{figure}

\begin{figure}[h]
 \includegraphics[scale=0.27]{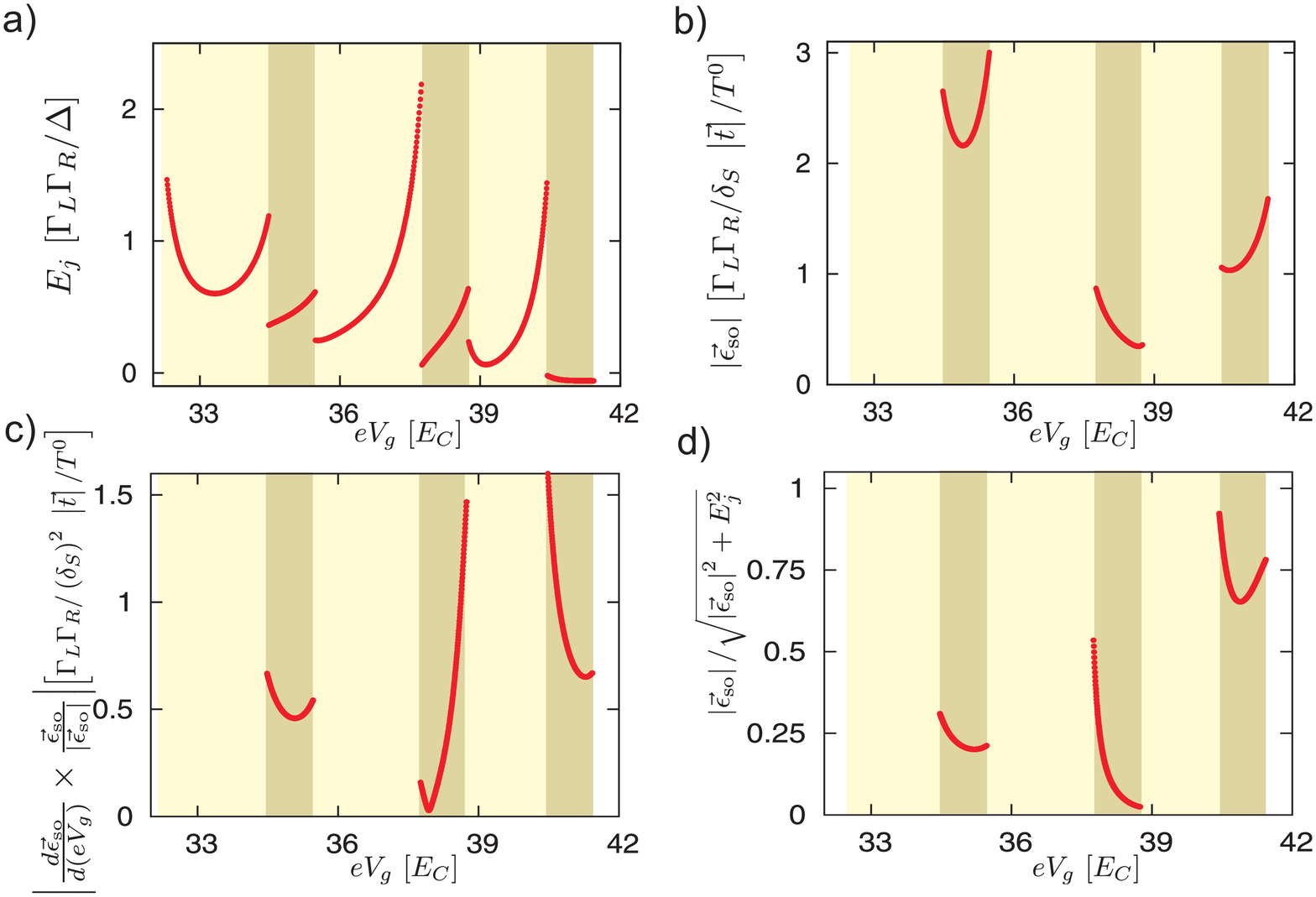}
 \caption{ The same as in Figure \ref{f3}, for the regime: $E_C/\Delta=1.5$, $\delta_S/E_C=1.5$
 \label{f5}}
 \end{figure}

\begin{figure}[h]
 \includegraphics[scale=0.27]{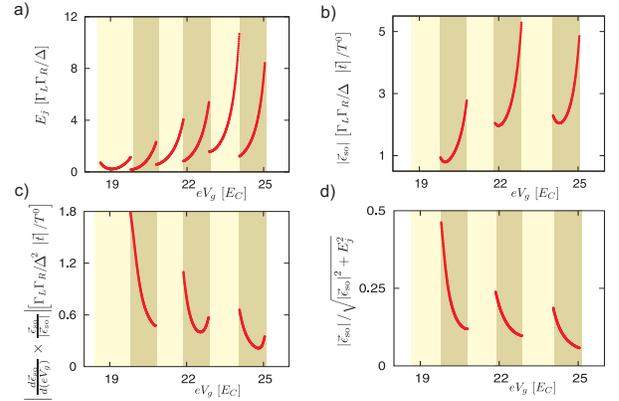}
 \caption{ The same as in Figure \ref{f3}, for the regime: $E_C/\Delta=1.5$, $\delta_S/E_C=0.1$
 \label{f6}}
 \end{figure}

We vary the gate voltage over a large domain, permitting observation of multiple Coulomb diamonds (see Figures \ref{f3}-\ref{f6}). The size of the diamonds observed is $E_C$ in the case of odd number of electrons and increases by the level spacing in the case of diamonds with even number of electrons.

The units in Figures \ref{f3}-\ref{f6} are chosen in accordance with the estimations presented in the section above, such that the value of unity corresponds to the estimated value of the quantity close to the edge, i.e. at $E_{\pm}=\Delta$. The results confirm our estimations.

We focus on four regimes differentiated by high charging effects $E_C/\Delta=10$, see Figures \ref{f3} and \ref{f4}, and relatively low charging effects $E_C/\Delta=1.5$, see Figures \ref{f5} and \ref{f6}. We also compare large average level spacing $\delta_S/E_C=1.5$, see Figures \ref{f3} and \ref{f5}, to regimes where the level spacing is smaller $\delta_S/E_C=0.1$, see Figures \ref{f4} and \ref{f6}. 

A general feature of the results in Figures \ref{f3}-\ref{f6} is that both spin-independent and spin-dependent terms in the Josephson energy increase as we approach edges of the diamond. Furthermore, comparing Figure \ref{f5}a, where $E_C\gtrsim \Delta$, with Figure \ref{f3}a, where $E_C\gg \Delta$, we can conclude that for lower charging energies the ratio between Josephson energy in the middle of diamond and energy at the edge is reduced, in agreement with our estimations. In Figures \ref{f5}-\ref{f6}a $E_j$ is shown to decrease in modulus towards the edge. This an unusual behavior that can be explained if there is a point in the higher energy state of the bistable region $E_{\pm}<\Delta$ where $E_j$ would change sign. 

Let us focus on the spin-independent contribution to the Josephson energy, presented in panel A of Figures \ref{f3}-\ref{f6}. In the single level regime, we expect $E_j$ to be positive for diamonds with even number of electrons and negative for odd occupancy, as explained in \cite{Dam2006}. This behavior, observed in Figure \ref{f3}a in the first two diamonds, would dominate for $\delta_S/E_C\gg 1$. In the regimes presented, an interesting feature occurs: there are regions of gate voltage where $E_j$ changes sign within a single diamond (point of supercurrent reversal \cite{Dam2006}). As a result, $E_j$ is dramatically suppressed. In these regions, $\left|\vec{\epsilon}_{\rm so}\right|$ may provide the dominating contribution to the total Josephson energy. This can be observed in panel D when $\left|\vec{\epsilon}_{\rm so}\right|/\sqrt{E_j^2+\left|\vec{\epsilon}_{\rm so}\right|^2}\approx 1$, see Figures \ref{f3} and \ref{f4}. Comparing the different regimes, we observe that regions where $E_j$ changes sign are more likely to appear if $\delta_S\approx E_C$. In addition, the larger variations of the Josephson energy observed for $E_C\gg\Delta$ compared to the regime $E_C\gtrsim \Delta$, further increase the probability of sign reversal. 

Turning to the polarization vector $\vec{\epsilon}_{\rm so}$, panels B and C of Figures \ref{f3}-\ref{f6} present its modulus and the variation of its orientation as a function of gate voltage. The modulus $\left|\vec{\epsilon}_{\rm so}\right|$ exhibits similar behavior as the absolute value of the spin-independent term $E_j$. In contrast to $E_j$, we do not observe regions of gate voltage where $\left|\vec{\epsilon}_{\rm so}\right|$ vanishes. It is not surprising: the probability that all three components of the polarization vector would vanish at the same value of gate voltage is very small.

In panel D of Figures \ref{f3}-\ref{f6} we plot the normalized ratio between the spin-dependent and spin-independent terms $\left|\vec{\epsilon}_{\rm so}\right|/\sqrt{E_j^2+\left|\vec{\epsilon}_{\rm so}\right|^2}$. We note that the ratio increases in the vicinity of regions where $E_j$ vanishes, as is the case in Figures \ref{f5}-\ref{f6}d. These regions are exceptional. The common case is represented by diamonds such as diamond 1 in Figure \ref{f3}d, where one observes a maximum of the ratio in the middle of the Coulomb diamond and decrease toward the edges. This is in agreement with our estimations, reflecting that $E_j$ diverges faster than $\left|\vec{\epsilon}_{\rm so}\right|$ as we approach the diamond edge.

Let us focus on the direction of $\vec{\epsilon}_{\rm so}$ and its dependence on gate voltage. We study the derivative $d\vec{\epsilon}_{\rm so}/dV_g$, projected in the perpendicular direction to $\vec{\epsilon}_{\rm so}$. As a general feature, the derivative is significantly reduced in the middle of the diamonds, as compared to the edge. 
It is interesting to note that the change in the perpendicular projection is not as smooth as the variation of $\vec{\epsilon}_{\rm so}$ with gate voltage. In the regime $\Delta\approx\delta_S$, see Figures \ref{f3}-\ref{f5}c, we find regions of gate voltage where the perpendicular derivative either vanishes, or abruptly changes its behavior. These features can be explained as follows. In this regime $N_S\approx 1$ near the edges, meaning that the direction of $\vec{\epsilon}_{\rm so}$ is dominated by contributions of few levels. It is possible to find regions of gate voltage where the weight of the dominating contribution vanishes, similar to the case already discussed for $E_j$. In the vicinity of such points, the weight of the dominating contribution changes sign. This is represented by the sharp turning points observed in Figures \ref{f3}-\ref{f5}c. In comparison, the case $\Delta\gg\delta_S$ presented in Figure \ref{f6}c shows that the behavior is smooth. In this regime the contribution is dominated by terms of multiple levels, reducing the probability to encounter values of the gate voltage where the dominating contribution to the perpendicular derivative vanishes. 

We may also conclude that $\vec{\epsilon}_{\rm so}$ varies mainly laterally when the gate voltage is set to the middle of the diamond. Here, the variation of $\left|\vec{\epsilon}_{\rm so}\right|$ vanishes. As the gate voltage is tuned towards the diamond edges, the lateral variation is overcome by the faster divergence of $\left|\vec{\epsilon}_{\rm so}\right|$. 

\section{Conclusion}
\label{conc}

In conclusion, we have outlined our proposal of spin superconducting qubit. Such unit would combine the natural representation of two level system in terms of electron spin and advantages of superconducting qubits. Spin and superconducting qubits can be operated within the circuit, the flux and spin degrees of freedom can be easily entangled. We have demonstrated feasibility of all electric manipulation of superconducting qubits and more complicated quantum gates made of such qubits.

The microscopic analysis presented shows that the spin-dependent part of the Josephson energy can be made sufficiently large, at least for semiconducting devices where spin-orbit interaction is intrinsically strong. We predict spontaneous breaking of time-reversal symmetry in the loops containing spin superconducting qubits. 

We acknowledge fruitful discussions with L. P. Kouwenhoven, H. Keijzers, and S. Frolov. This work is part of the research program of the Stichting FOM.

\end{document}